\documentclass[aps,prf,a4paper,groupedaddress,floatfix,showpacs,longbibliography]{revtex4-1}
\usepackage{t1enc}
\usepackage{amsmath,amssymb,mathtools,epstopdf,epsfig,subfigure}
\usepackage{hhline,bbm,psfrag}
\usepackage{subfigure}
\usepackage{graphicx}

\newcommand{\pd}{\partial}
\newcommand{\dt}{{\Delta t}}

\newcommand{\lap}{\nabla^2}

\newcommand{\divv}{\nabla\cdot}
\newcommand{\be}{\mathbf{e}}
\newcommand{\gap}{d}

\newcommand{\bU}{\mathbf{U}}
\newcommand{\bUb}{\mathbf{\overline{U}}}
\newcommand{\bu}{\mathbf{u}}

\newcommand{\mL}{\mathcal{L}}

\newcommand{\mN}{\mathcal{N}}

\newcommand{\rin}{r_{\rm in}}
\newcommand{\rout}{r_{\rm out}}
\newcommand{\Rin}{Re_{\rm in}}
\newcommand{\Routinit}{Re_{\rm out}}
\newcommand{\Rout}{Re}
\begin{document}
\title{Spirals and ribbons in counter-rotating Taylor-Couette flow: frequencies from mean flows and heteroclinic orbits}
\author{Yacine Bengana}
\affiliation{Laboratoire de Physique et M\'ecanique des Milieux 
H\'et\'erog\`enes (PMMH),
CNRS, ESPCI Paris, PSL Research University; Sorbonne Universit\'e, 
Univ. Paris Diderot, 75005 Paris France}
\author{Laurette S. Tuckerman}
\affiliation{Laboratoire de Physique et M\'ecanique des Milieux 
H\'et\'erog\`enes (PMMH),
CNRS, ESPCI Paris, PSL Research University; Sorbonne Universit\'e, 
Univ. Paris Diderot, 75005 Paris France} \email{laurette@pmmh.espci.fr}
\author{}
\affiliation{~~~\\{\it Physical Review Fluids} {\bf 4}, 044402 (2019)}


\begin{abstract}
  A number of time-periodic flows have been found to have a property
  called RZIF: when a linear stability analysis is carried out about
  the temporal mean (rather than the usual steady state), an
  eigenvalue is obtained whose {\bf Real} part is {\bf Zero} and whose
  {\bf Imaginary} part is the nonlinear {\bf Frequency}.  For
  two-dimensional thermosolutal convection, a Hopf bifurcation leads
  to traveling waves which satisfy the RZIF property and standing
  waves which do not. We have investigated this property numerically
  for counter-rotating Couette-Taylor flow, in which a Hopf
  bifurcation gives rise to branches of upwards and downwards
  traveling spirals and ribbons which are an equal superposition of
  the two. In the regime that we have studied, we find that both spirals
  and ribbons satisfy the RZIF property.
  As the outer Reynolds number is increased, the ribbon branch is succeeded by
  two types of heteroclinic orbits, both of which connect saddle states 
  containing two axially stacked pairs of axisymmetric vortices.
  One heteroclinic orbit is non-axisymmetric, with excursions
  that resemble the ribbons, while the other remains axisymmetric.
\end{abstract}

\maketitle

\section{Introduction}

A number of time-periodic flows have been found to have the following
property: when a linear stability analysis is carried out about the
temporal mean (rather than the usual steady state from which they
bifurcate), an eigenvalue emerges whose imaginary part reproduces the
frequency of the nonlinear oscillation and whose real part is near
zero.  This property was first discovered by Barkley
\cite{barkley2006linear} and extensively studied
\cite{pier2002frequency,noack2003hierarchy,mittal2008global,mantivc2014self,mantivc2015self,sipp2007global}
for the cylinder wake (for which it is strongly verified)
and then for an open cavity
\cite{sipp2007global,meliga2017harmonics,bengana2018} (for which it holds
quite well, though not exactly). In these two-dimensional
flows, a horizontal velocity imposed at infinity is deviated as it
encounters objects (a cylinder) or openings (a cavity) and, for
sufficiently high Reynolds number, undergoes a Hopf bifurcation.

In Turton, Tuckerman \& Barkley \cite{turton2015prediction}, this
property was named RZIF -- a mnemonic for the fact that the eigenvalue
obtained by linearizing about the mean flow has a {\bf Real} part of
{\bf Zero} and an {\bf Imaginary} part which is the nonlinear {\bf
  Frequency} -- and was studied for a quite different kind of flow.
Thermosolutal convection is driven by a vertical density gradient in a
fluid mixture between two horizontal plates, which in turn is caused
by imposed temperature and concentration differences at the two
bounding plates. When the temperature and concentration have opposing
effects on the density, instability is manifested as a Hopf
bifurcation.  In a two-dimensional domain with horizontally periodic
boundary conditions, this is a classic case of the breaking
of $O(2)$ symmetry (translations and reflection) and
leads to the simultaneous formation of traveling and standing
wave branches, at most one of which can be stable
\cite{knobloch1986oscillatory}.  Turton et
al.~\cite{turton2015prediction} discovered that the traveling waves in
thermosolutal convection provide another clear realisation of the RZIF
property, while the standing waves provide an equally clear
counterexample: the real part of the eigenvalue obtained by
linearizing about the mean part of the standing waves is far from zero
and its imaginary part is far from the nonlinear frequency.

\begin{figure}[t]
\centerline{\includegraphics[width=0.9\columnwidth]{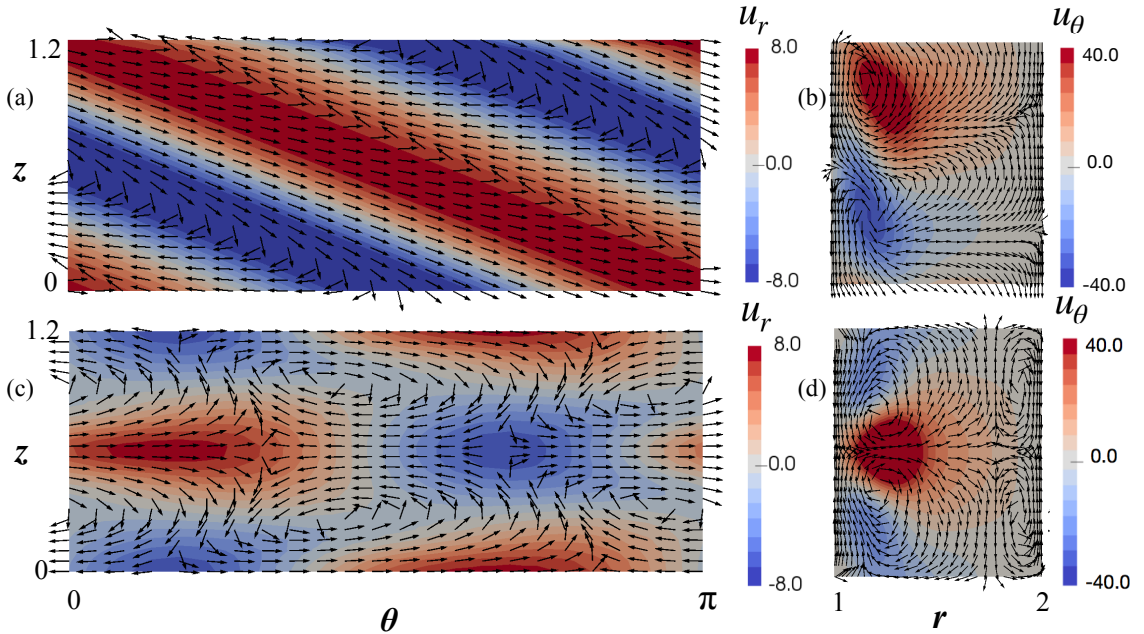}}
  \caption{Nonlinear spirals (a,b) and ribbons (c,d) at $\Rout=-550$.
    The deviation from laminar Couette flow is shown.
    Spirals of the opposite helicity exist and have identical properties but are not shown.
    a,c) $(\theta,z)$ velocity (arrows) with $u_r$ (colors) at $r=1.26$,
    near the maximum of the deviation from laminar Couette flow.
    b,d) $(r,z)$ velocity (arrows) with $u_\theta$ (colors).}
\label{fig:nonlinearvisu}
\end{figure}

We have sought to continue this investigation on another pair of
hydrodynamic traveling and standing waves.  In Taylor-Couette flow,
when only the inner cylinder rotates, a steady bifurcation leads to the
classic Taylor vortices. In counter-rotating Taylor-Couette flow,
however, these vortices are inclined to the rotation axis, and hence
non-axisymmetric. A Hopf bifurcation then gives rise to branches of
the well-known spirals
\cite{krueger1966relative,langford1988primary,antonijoan1998non} and
the somewhat lesser-known ribbons
\cite{demay1984calcul,tagg1989nonlinear,chossat2012couette,pinter2006competition,pinter2008bifurcation,pinter2008wave,hoffmann2009transitions,altmeyer2010secondary,deguchi2013fully}, which,
like standing waves, are an equal superposition of spirals moving
axially upwards and downwards.
As in \cite{knobloch1986oscillatory,turton2015prediction},
this is a consequence of the breaking of $O(2)$ symmetry,
here in the axial direction, when end effects are disregarded.
Figure \ref{fig:nonlinearvisu} shows the appearance of these two
types of flows.

Counter-rotating Taylor-Couette flow has an additional symmetry
when compared to the two-dimensional thermosolutal cases 
studied in \cite{knobloch1986oscillatory,turton2015prediction}.
Along with the $O(2)$ axial $z$ symmetry, the Taylor-Couette
configuration has $SO(2)$ symmetry, 
consisting of rotations (not reflection) in the azimuthal $\theta$ direction.
The tilted spirals or azimuthally tiled ribbons are not axisymmetric,
and hence they break $SO(2)$ symmetry.
In such cases, bifurcations always lead to states which rotate in
the azimuthal direction; see e.g. \cite{ecke1992hopf}.
Thus, although ribbons are standing waves in the axial direction,
i.e. equal superpositions of axially-upwards and axially-downwards
traveling waves, they rotate in the azimuthal direction,
as do the spirals.

As the rotation velocity of the outer cylinder is varied, the ribbon
branch is succeeded by two types of heteroclinic orbits, both of which
connect saddle states consisting of two pairs of axisymmetric
vortices.  One heteroclinic orbit is non-axisymmetric, with excursions
that resemble the ribbons, while the other remains axisymmetric.

\section{Methods}

To investigate this problem numerically, 
we have used the spectral finite-difference code of Willis
\cite{dessup2018}
to solve the Navier-Stokes equations in a
cylindrical annulus, 
\begin{subequations}
\begin{align}
\pd_t \bU &= -\left(\bU\cdot\nabla\right)\bU -\nabla P + \lap\bU \\
  \divv \bU &= 0 \\
  \bU &= \be_\theta \Rin, \be_\theta\Routinit \quad\mbox{~at~} \rin, \rout
\end{align}
\label{eq:nonlin}\end{subequations}
where the length scale is $\gap\equiv\rout-\rin$ and the
time scale is $\gap^2/\nu$, where $\nu$ is the kinematic viscosity.
The radius ratio is $\eta\equiv \rin/\rout$, 
the two Reynolds numbers are defined as 
$Re_j\equiv r_j\Omega_j\gap/\nu$ where $j={\rm in}$, out, 
and the numerical resolution is $(N_r,N_\theta,N_z)=(33,16,16)$.

We use the parameters in Pinter et al.~\cite{pinter2006competition,pinter2008bifurcation,pinter2008wave}, 
namely radius ratio $\eta\equiv \rin/\rout=0.5$, inner Reynolds number $\Rin=240$ and
azimuthal wavenumber $M_0=2$ (azimuthal wavelength $2\pi/M_0=\pi$).
We have chosen the axial wavelength $\lambda_z=1.2$, slightly less than the value 
studied intensively by \cite{pinter2006competition,pinter2008bifurcation,pinter2008wave}
to avoid a secondary bifurcation they describe near the onset of spirals and ribbons for 
$\lambda_z=1.3$.
Because counter-rotating Taylor-Couette flow is centrifugally unstable 
only near the inner cylinder, here for $r\in[1,1.24]$, the preferred axial 
wavelength is correspondingly less than that which would be expected if the 
vortices filled the gap.
Because $\Rin$ remains fixed and we will vary $\Routinit$, we will simplify
the notation by using $\Rout$ instead of $\Routinit$ to designate the outer Reynolds number.

We have modified the Taylor-Couette code \cite{dessup2018} in order to timestep 
the equations linearized about a velocity field $\bU$:
\begin{subequations}
\begin{align}
\pd_t \bu &= -\left(\bU\cdot\nabla\right)\bu - \left(\bu\cdot\nabla\right) \bU 
-\nabla p + \lap\bu \label{eq:linstaba}\\
  \divv \bu &= 0 \label{eq:linstabb}\\
  \bu &= 0 \quad\mbox{~at~} \rin, \rout \label{eq:linstabc}
\end{align}
\label{eq:linstab}\end{subequations}
Time-integration of the linearized equations 
\eqref{eq:linstab} about a steady solution $\bU$ converges to the leading 
eigenvector, i.e. that whose corresponding eigenvalue 
has the most positive or least negative real part or growth rate.
It is a means of carrying out the power method on the
exponential of the right-hand-side of \eqref{eq:linstab}.
Integration of \eqref{eq:linstab} does not generally lead to
a steady or periodic state, since the nonlinear terms which
would otherwise saturate the amplitude is absent.
Instead, it leads to states whose amplitude 
increases or decreases, but with a constant growth rate
and frequency, in the case of a complex eigenvalue.

We first use the linearized code to compute the Hopf bifurcation 
threshold from laminar Couette flow.
For various timesteps $\dt$, 
we calculate the growth rate $\sigma$ as a function of $\Rout$
and then use linear extrapolation or interpolation to find $Re_c$ 
such that $\sigma(Re_c)=0$. (This procedure circumvents critical slowing
down and so finds bifurcation thresholds faster and
more accurately than could be done with nonlinear simulations.)
The bifurcation thresholds calculated using
$\dt=10^{-4}$, $10^{-5}$, and $10^{-6}$ are, respectively,
$Re_c = -595.23$, $Re_c = -586.65$, $Re_c = -585.42$, the last of which we
consider to be the converged value. (For greater legibility and simplicity,
in what follows we cite its truncated value $-585$.)

The branches of spirals and ribbons illustrated in figure \ref{fig:nonlinearvisu}
bifurcate at $\Rout \approx -585$ towards increasing $\Rout$.
Their functional form is:
\begin{eqnarray}
  u^{{\rm SPI}\pm}(r,\theta,z,t) &=& \sum_{m}\hat{\bu}_{m}(r)\:e^{im (z/\lambda_z \pm M_0\theta-\omega_{\rm SPI} t)}\nonumber\\
  u^{\rm RIB}(r,\theta,z,t) &=& \sum_{k,m}
 \left\{\begin{array}{c}\hat{u}_{km,r}(r)\be_r\:\sin(kz/\lambda_z)\\
\hat{u}_{km,\theta}(r)\be_\theta\:\sin(kz/\lambda_z)\\
\hat{u}_{km,z}(r)\be_z\:\cos(kz/\lambda_z)\end{array}\right\}
\:e^{im(M_0\theta-\omega_{\rm RIB} t)}
                                \label{eq:funcform}
\end{eqnarray}
The spirals are traveling waves: they depend on $z$, $\theta$, and $t$
only via the single combination
$z/\lambda_z+ M_0\theta-\omega_{\rm SPI}t$.  The ribbons have fixed
axial nodal lines and are reflection symmetric; 
the axis of symmetry is $z=\lambda_z/2$ in equation \eqref{eq:funcform} 
and in figure \ref{fig:nonlinearvisu}. 
The ribbons are traveling waves in $\theta$:
they depend on $\theta$ and $t$ via $M_0\theta-\omega_{\rm RIB}t$.
Many other complex and esthetic symmetry-breaking flows have been
observed in numerical simulations of this configuration
\cite{pinter2006competition,pinter2008wave,pinter2008bifurcation,hoffmann2009transitions,altmeyer2010secondary,deguchi2013fully},
including cross spirals, oscillating cross spirals, mixed cross
spirals and mixed ribbons.

We will use numerical simulations of the nonlinear and linearized
equations \eqref{eq:nonlin} and \eqref{eq:linstab} to investigate the RZIF
phenomenon mentioned in the introduction, in which linearization about
the mean flow of a periodic solution yields an eigenvalue whose {\bf
  Real} part is {\bf Zero} and whose {\bf Imaginary} part is the
nonlinear {\bf Frequency}.  To do so, we will calculate the spirals
and ribbons by time integrating the nonlinear equations \eqref{eq:nonlin} until an asymptotic
state is reached.  Because the spirals and ribbons are both
azimuthally traveling waves, their temporal means are identical with
their azimuthal means which are, in turn, the axisymmetric $(m=0)$
components of \eqref{eq:funcform}.  We will then find the eigenvalues
and eigenvectors about the resulting mean flows by integrating
equations \eqref{eq:linstab}.

\section{RZIF analysis}

\begin{figure}[t]
\centerline{\includegraphics[width=0.9\columnwidth]{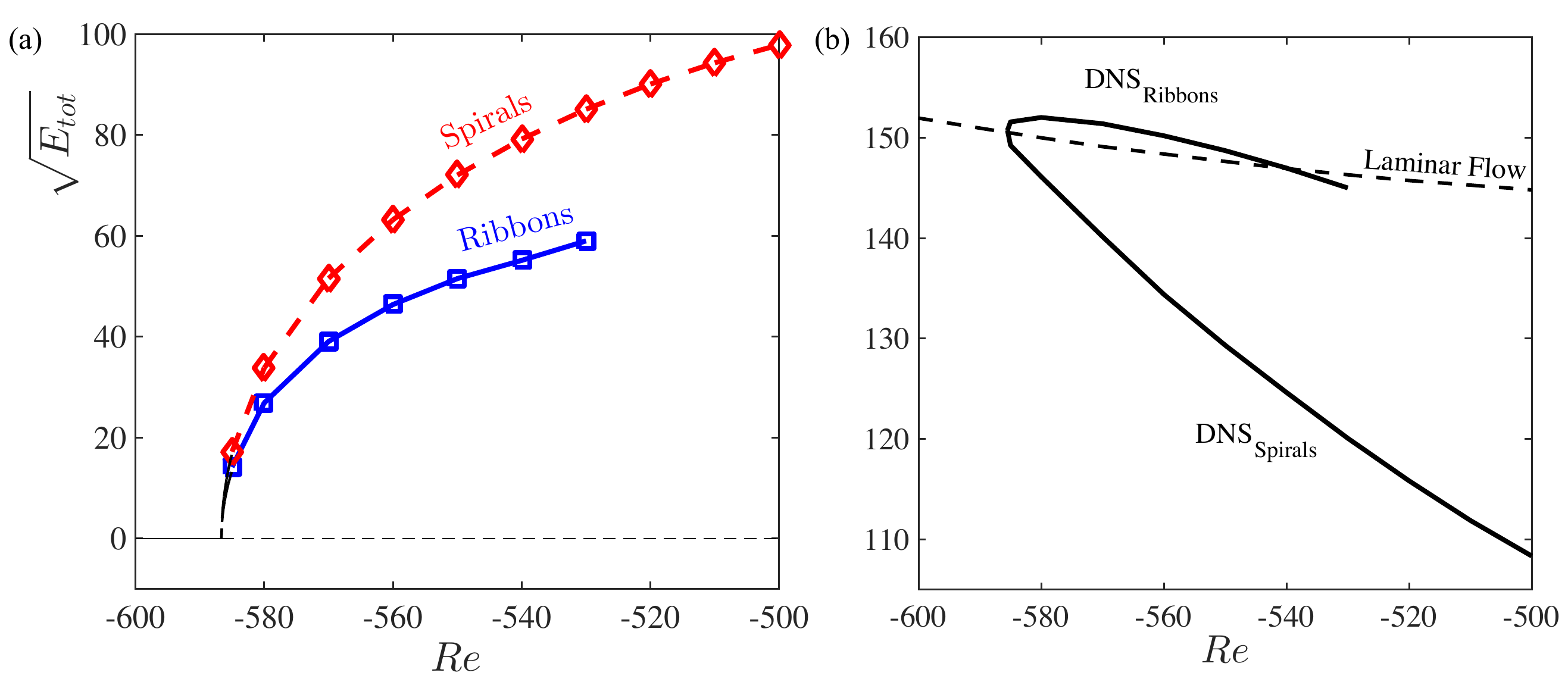}}
  \caption{a) Bifurcation diagram
    for spirals and ribbons. Square root of kinetic energy of the deviation from
    laminar Couette flow is shown as a function of $\Rout$.
    The ribbons are stable where shown, while the spirals are
    calculated by using a spiral initial condition.
    b) Frequencies of spiral and ribbon states as a function of $\Rout$.
    The dashed line shows the eigenvalues obtained by linearizing about laminar Couette flow.}
\label{fig:bifdiag}
  \end{figure}
\begin{figure}
\centerline{
  \includegraphics[width=0.9\columnwidth]{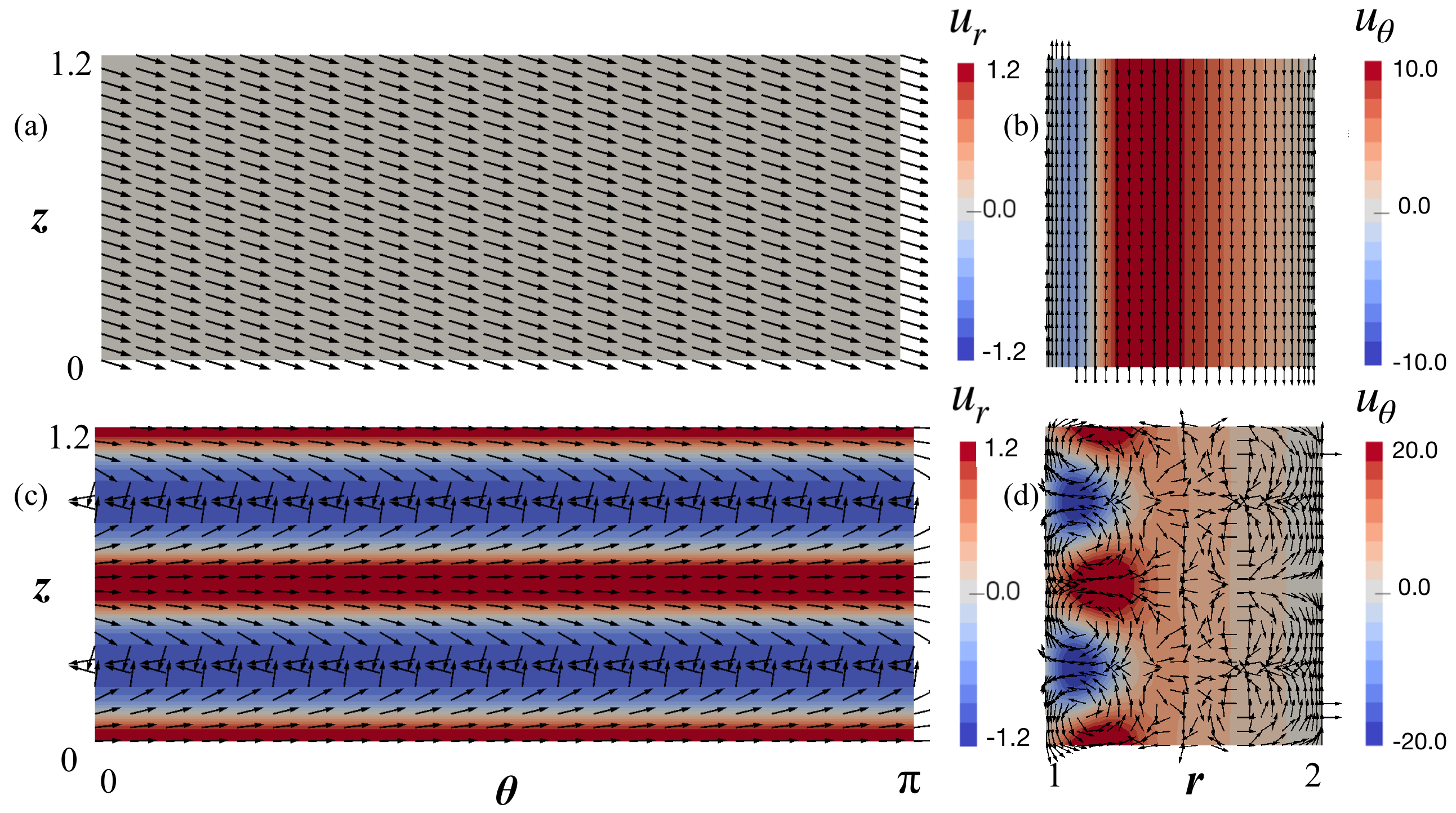}}
  \caption{Mean flow of spirals (a,b) and ribbons (c,d) at $\Rout=-550$.
    The deviation from laminar Couette flow is shown.
    a,c) $(\theta,z)$ velocity (arrows) with $u_r$ (colors) at $r=1.26$,
    near the maximum of the deviation from laminar Couette flow.
    b,d) $(r,z)$ velocity (arrows) with $u_\theta$ (colors).
    Since the spirals are traveling waves in the $\theta$ and $z$ directions,
    their means have no dependence on either. The ribbons are traveling waves
    in $\theta$ and hence their mean is axisymmetric.}
\label{fig:meanvisu}
\vspace*{1cm}
\centerline{
  \includegraphics[width=0.85\columnwidth]{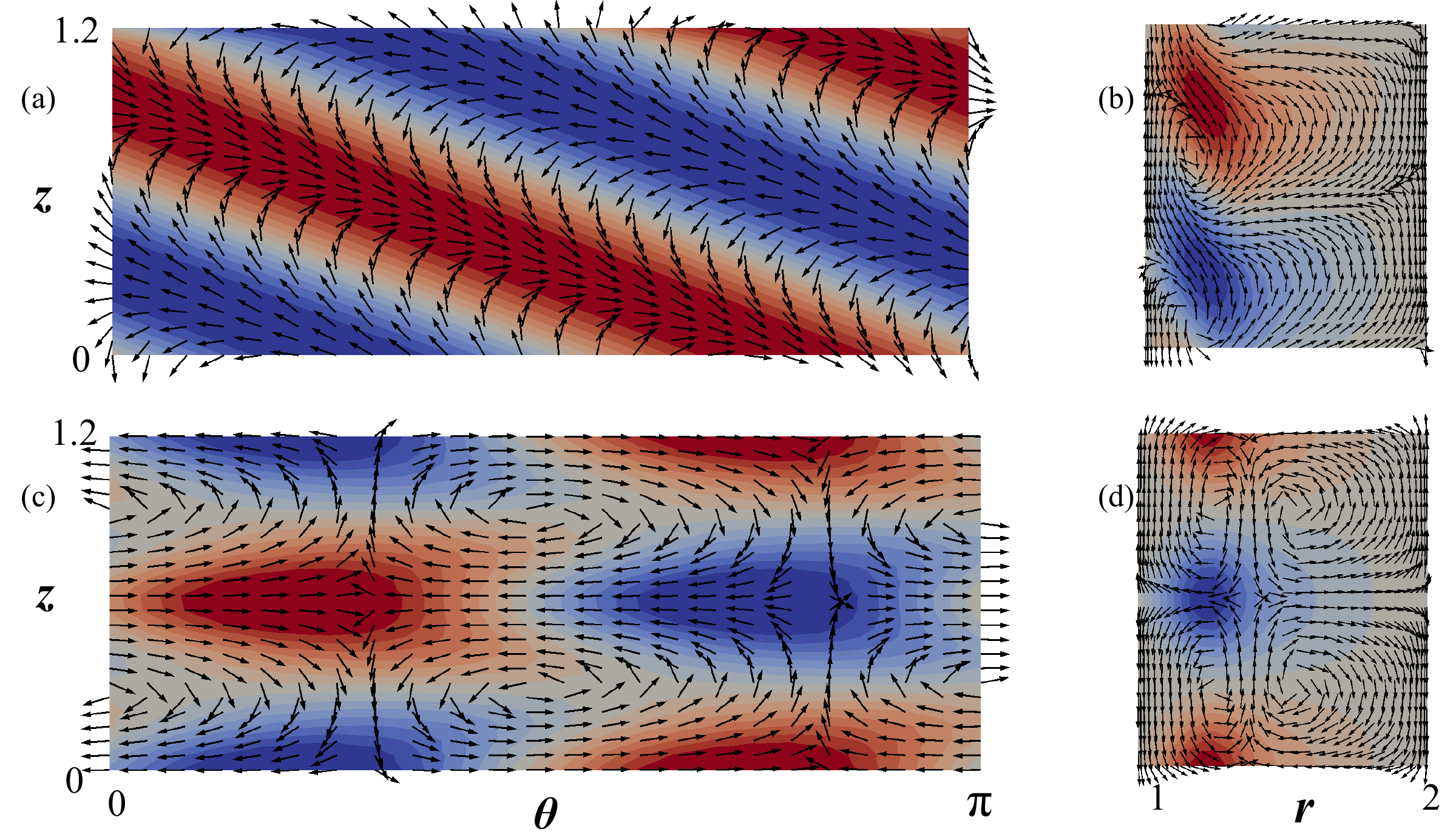}}
\caption{Eigenvectors resulting from linearization about mean flow of
  spirals (a,b) and ribbons (c,d) at $\Rout=-550$.  a)
  $(\theta,z)$ velocity (arrows) with $u_r$ (colors) at
  $r=1.26$.  b) $(r,z)$ velocity (arrows) with
  $u_\theta$ (colors).  These eigenvectors resemble the deviation of
  nonlinear spirals and ribbons from the laminar Couette flow as well
  as the eigenvectors of spiral and ribbon form obtained from
  linearizing about the laminar flow.}
\label{fig:evecvisu}
  \end{figure}
\begin{figure}
\centerline{
  \includegraphics[width=0.85\columnwidth]{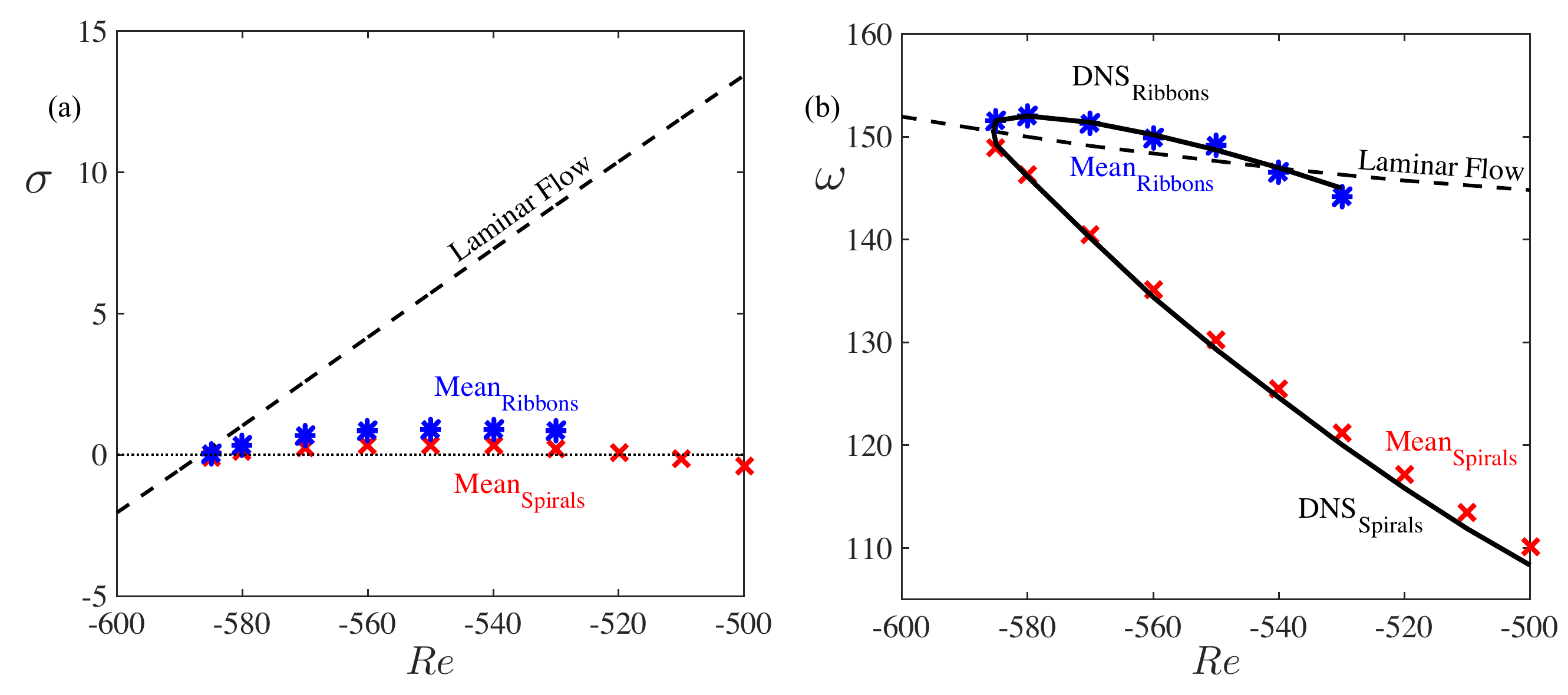}}
  \caption{Growth rates (a) and frequencies (b).  Results from
    nonlinear numerical simulations (black curves) and
    from Navier-Stokes equations linearized about the laminar flow
    (dashed lines), about the mean flow of spirals (red crosses), and
    about the mean flow of ribbons (blue stars).  Satisfaction of the RZIF
    property for the frequencies is demonstrated by the agreement
    between the red crosses and the black curve for spirals and
    between the blue stars and the black curve for ribbons.
    Satisfaction of the RZIF property for the growth rates is
    demonstrated by the fact that the blue stars and the red crosses
    are near zero.}  
  \label{fig:RZIF}
  \end{figure}

\begin{figure}
\centerline{
  \includegraphics[width=0.85\columnwidth]{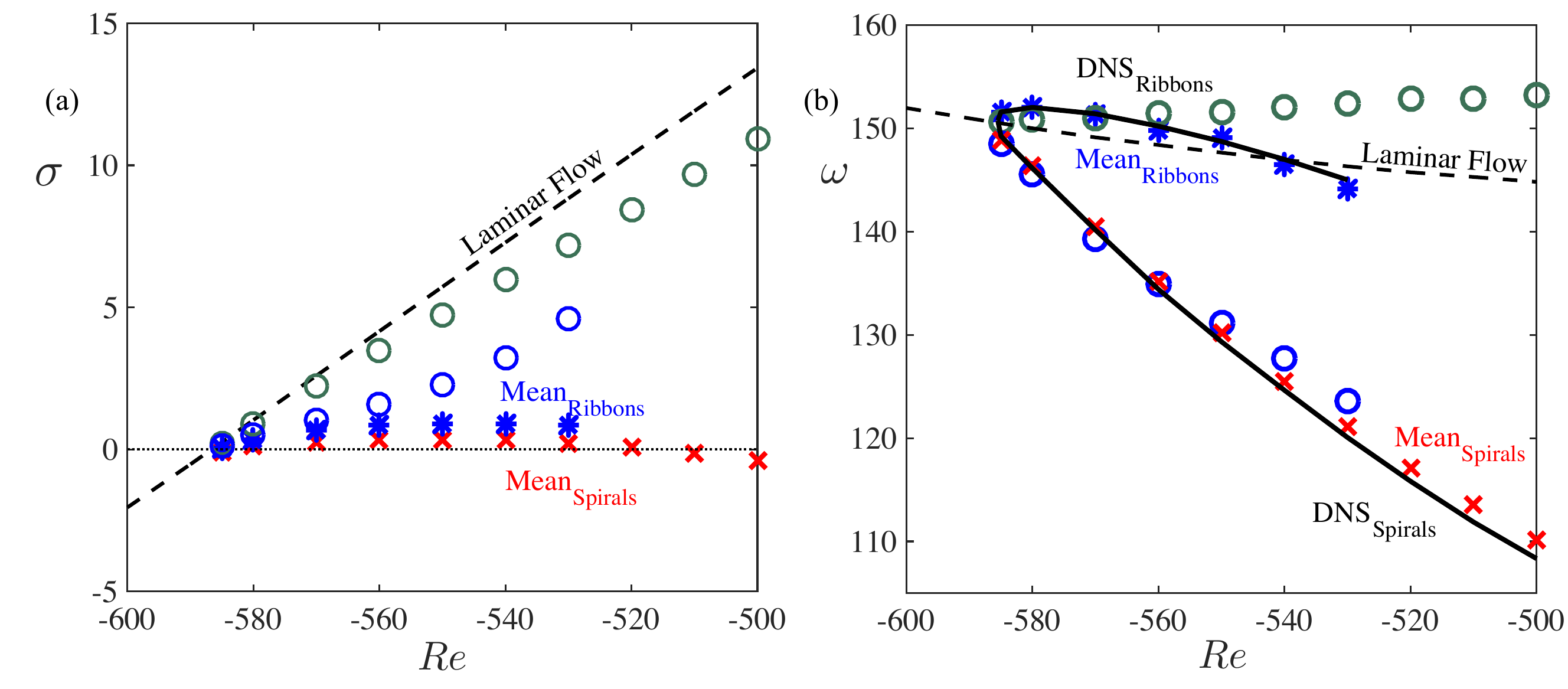}}
  \caption{Same as figure \ref{fig:RZIF} with additional eigenvalues.
   The hollow circles are eigenvalues corresponding
    to spiral eigenvectors.  The dark green circles (upper set) are obtained by
    linearizing around a mean spiral flow whose orientation is
    opposite to that of the eigenvector while the blue ones (lower set) are
    obtained by linearizing around the mean flow of a ribbon.}
  \label{fig:RZIF_other}
  \end{figure}
\begin{figure}[t]
\centerline{
  \includegraphics[width=0.85\columnwidth]{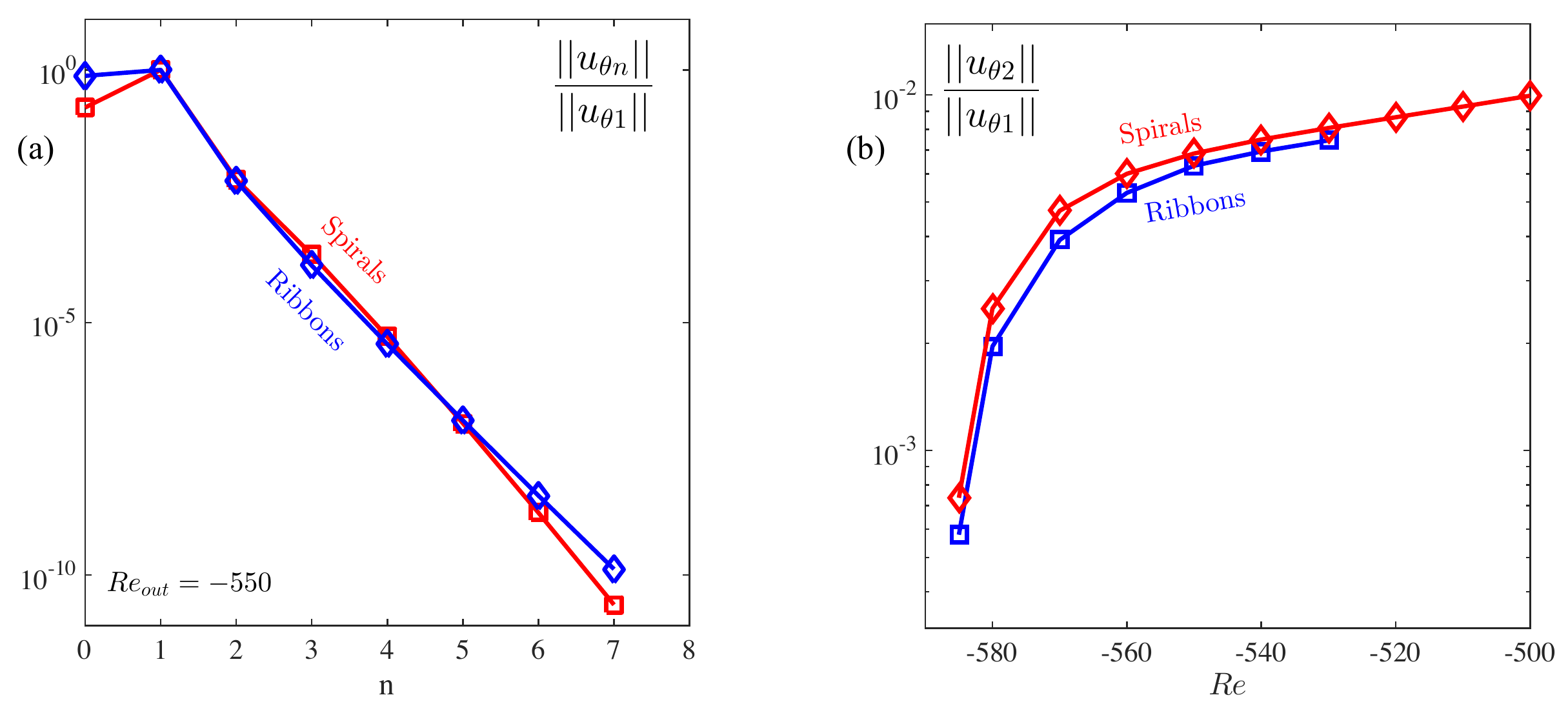}}
  \caption{Spectra of spirals and ribbon states.
    a) Full spectrum for states at $\Rout=-550$, normalized by amplitude of fundamental frequency. 
    b) Ratio of second harmonic to fundamental frequency.}
  \label{fig:spectra}
  \end{figure}

We begin by showing the bifurcation diagram for spirals and ribbons.
Their common threshold is $\Rout \approx -585$.  On figure
\ref{fig:bifdiag}a, we represent the square root of kinetic energy of
the two branches.  In this parameter range, the spirals are unstable
to ribbons, but can be computed by including only $k=m$ modes or, like
\cite{pinter2006competition}, by setting the $m=0$ component of $u_r$
to zero.  The ribbons themselves become unstable for
$\Rout \gtrsim -530$.  Figure \ref{fig:bifdiag}b shows the frequency
of the nonlinear spirals and ribbons, as well as those
obtained by linearizing about laminar Couette flow (dashed line).
Linearization about the laminar flow leads to the same set of
eigenvalues for spirals or ribbons, since the eigenvectors
corresponding to upgoing and downgoing spirals are related by
symmetry, with their superposition leading to ribbons.  For a
supercritical Hopf bifurcation, these necessarily match the frequency
of the nonlinear branches at the threshold.  As $\Rout$ is increased,
the frequencies of the nonlinear spirals deviate greatly from those
derived from the laminar flow, but the frequencies of the nonlinear
ribbons remain close.  This was also true of the standing waves of
thermosolutal convection \cite{turton2015prediction}, which inspired
this study: the frequencies of the nonlinear standing waves were much
closer to those obtained by linearization about the laminar
(conductive) flow than were the frequencies of the nonlinear traveling
waves.

Figure \ref{fig:meanvisu} shows the temporal means of the nonlinear 
spiral and ribbon states shown in figure \ref{fig:nonlinearvisu}
at $\Rout=-550$.
Since spirals are traveling waves in both the $\theta$ and the $z$
direction, their temporal means are equivalent to their spatial means
and have no dependence on either direction.  Ribbons are traveling
waves in the $\theta$ direction and so their means have no $\theta$
dependence.
We set $\bU$ in the linearized equations \eqref{eq:linstab}
to the mean flows of spirals or ribbons
shown in figure \ref{fig:meanvisu} and initialize $\bu$
with fields of the spiral or ribbon form, respectively.
We then integrate until the growth rate and the frequency are constant.
The resulting field $\bu(t)$ then cycles between the real and imaginary
parts of the eigenvector which here are related by a spatial shift.
Figure \ref{fig:evecvisu} shows samples of these eigenvectors.
These greatly resemble the deviations of the
corresponding nonlinear flows from laminar Couette flow shown in 
figure \ref{fig:nonlinearvisu} and resemble even more the 
usual eigenvectors of spiral and ribbon form resulting from
linearization about laminar Couette flow.

Figure \ref{fig:RZIF} presents the main result of this section, namely
that both the spirals and the ribbons satisfy the RZIF property.
The growth rates or real parts of eigenvalues are shown in figure \ref{fig:RZIF}a
and the frequencies or imaginary parts are shown in figure \ref{fig:RZIF}b.
The solid curves show the frequencies obtained from 
nonlinear simulations.  Linearization about the mean flows of
spirals (red crosses) or of ribbons (blue stars) lead to eigenvalues
whose imaginary parts agree with the nonlinear frequencies.  The corresponding
growth rates are seen to be near
zero, i.e. the mean flows are nearly marginally stable. (This has no
relation to the marginal stability of the full spirals or ribbons to
shifted versions of themselves, which arises from their $\theta$ or
$z$ dependence in a homogeneous domain.) Thus, both the spirals and the
ribbons satisfy the RZIF property.
This contrasts with the thermosolutal case, in which 
the traveling waves satisfied the RZIF property, 
while the standing waves did not.

Additional eigenvalues are shown in figure \ref{fig:RZIF_other}, which
are not related to the basic RZIF property.  Both correspond to spiral
eigenvectors and have positive growth rates. These show the frequency
and growth rate of spiral eigenvectors superposed on the mean flow of
spirals of opposite helicity (dark green circles) or of ribbons (blue
circle).  The positive growth rate for the spirals of opposite
chirality could be related to the linear instability of the spirals to
ribbons; such a transition would take place via the addition of
spirals of opposite chirality until equal amplitudes are reached.
However, for these parameter values, ribbons show no tendency to 
make a transition to spirals, and yet the growth rates from 
the mean flow of ribbons to spirals are also positive (although 
less so than the growth rates from the mean flow of spirals).
More understanding of the meaning of linearization about mean
flows would be necessary to interpret these eigenvalues.

The RZIF property is implied by a near-monochromatic spectrum, 
as shown in \cite{turton2015prediction} as follows.
Consider the evolution equation
\begin{align}
\pd_t \bU &=\mL \bU + \mN(\bU,\bU)
\label{eq:sys}\end{align}
where $\mL$ is linear and $\mN(\cdot,\cdot)$ is a quadratic nonlinearity.
Let
\begin{align}
\bU = \bUb + \sum_{n\neq 0} \bu_n e^{in\omega t}
\label{eq:periodic}\end{align}
(with $\bu_{-n}=\bu_n^\ast$)
be the temporal Fourier decomposition of
a periodic solution to \eqref{eq:sys} with mean $\bUb$ and frequency $\omega$. 
The $n=1$ component of \eqref{eq:sys} is 
%
\begin{align}
i\omega \bu_1 &= \underbrace{\mL \bu_1 + \mN(\bUb,\bu_1) + \mN(\bu_1,\bUb)}_{\displaystyle\mL_{\bar{U}}\bu_1} + 
\underbrace{\mN(\bu_2,\bu_{-1}) + \mN(\bu_{-1},\bu_2) 
      + \mN(\bu_3,\bu_{-2}) + \mN(\bu_{-2},\bu_3) + \ldots}_{\displaystyle\mN_1}\label{eq:first}
        \end{align}
If, as is often the case, $||u_n||\sim \epsilon^{|n|}$, then
$\mN_1 = O(\epsilon^3)$ may be neglected and RZIF is satisfied:
the linear operator $\mL_{\bar{U}}$ in \eqref{eq:first}
has the pure imaginary eigenvalue $i\omega$,
corresponding to the frequency of the periodic solution.
Hence the RZIF property is satisfied for near-monochromatic oscillations 
in any system with a quadratic nonlinearity.

It remains to verify whether these periodic solutions are 
near-monochromatic.
Figure \ref{fig:spectra} shows the temporal Fourier spectra of 
the azimuthal velocity for the spirals and the ribbons.
For the thermosolutal case of
\cite{turton2015prediction}, the ratio $||u_2||/||u_1||$ was found to
be much higher for the standing waves (greater than about $10^{-1}$)
than for the traveling wave branch (between $10^{-3}$ and about
$10^{-1}$), explaining the satisfaction of RZIF for the traveling
waves and not for the standing waves.  In contrast, for this
counter-rotating Taylor-Couette case, figure \ref{fig:spectra} shows
that the temporal spectra for spirals and ribbons are quite similar.
The ratio $||u_2||/||u_1||$ is consistently less than $10^{-2}$ for
both flows over the range of our investigation.  This is consistent
with the fact that RZIF is satisfied for both flows.

\section{Heteroclinic orbits}

\begin{figure}
\centerline{\includegraphics[width=0.9\columnwidth]{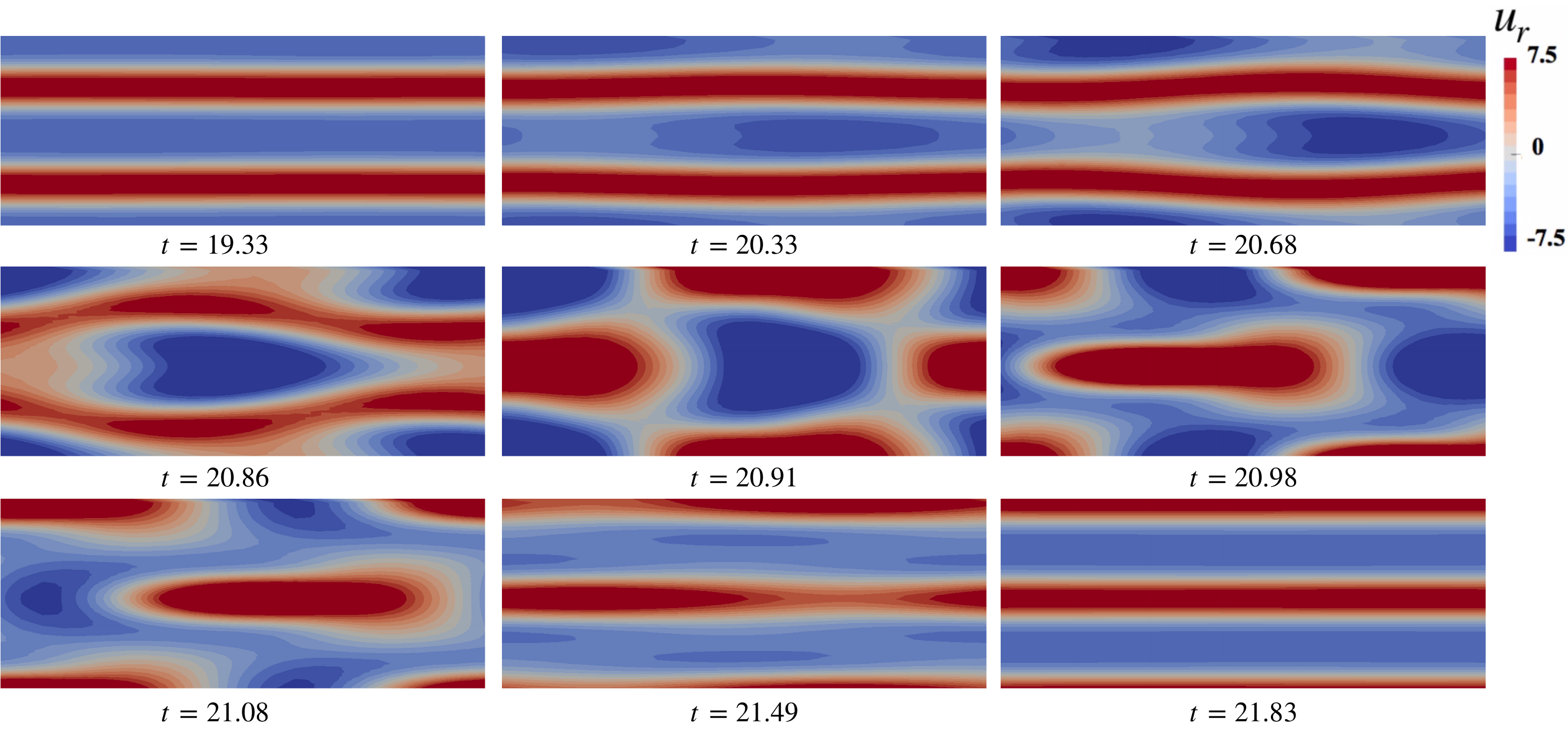}}
  \caption{Non-axisymmetric heteroclinic cycle at $\Rout=-502$.
Colors show the radial velocity in a $(\theta,z)$ slice at $r=1.26$.
The first (top left) and last (bottom right) visualizations show the
saddles which anchor the heteroclinic cycles. 
These contain two pairs of axisymmetric vortices and differ by an 
axial phase shift of a single vortex.
The visualizations between the two illustrate the excursion or rapid transition 
between the two saddles, via states which resemble ribbons.}
\label{fig:Het_Snapshots_502}
\end{figure}
\begin{figure}
\centerline{\includegraphics[width=\columnwidth]{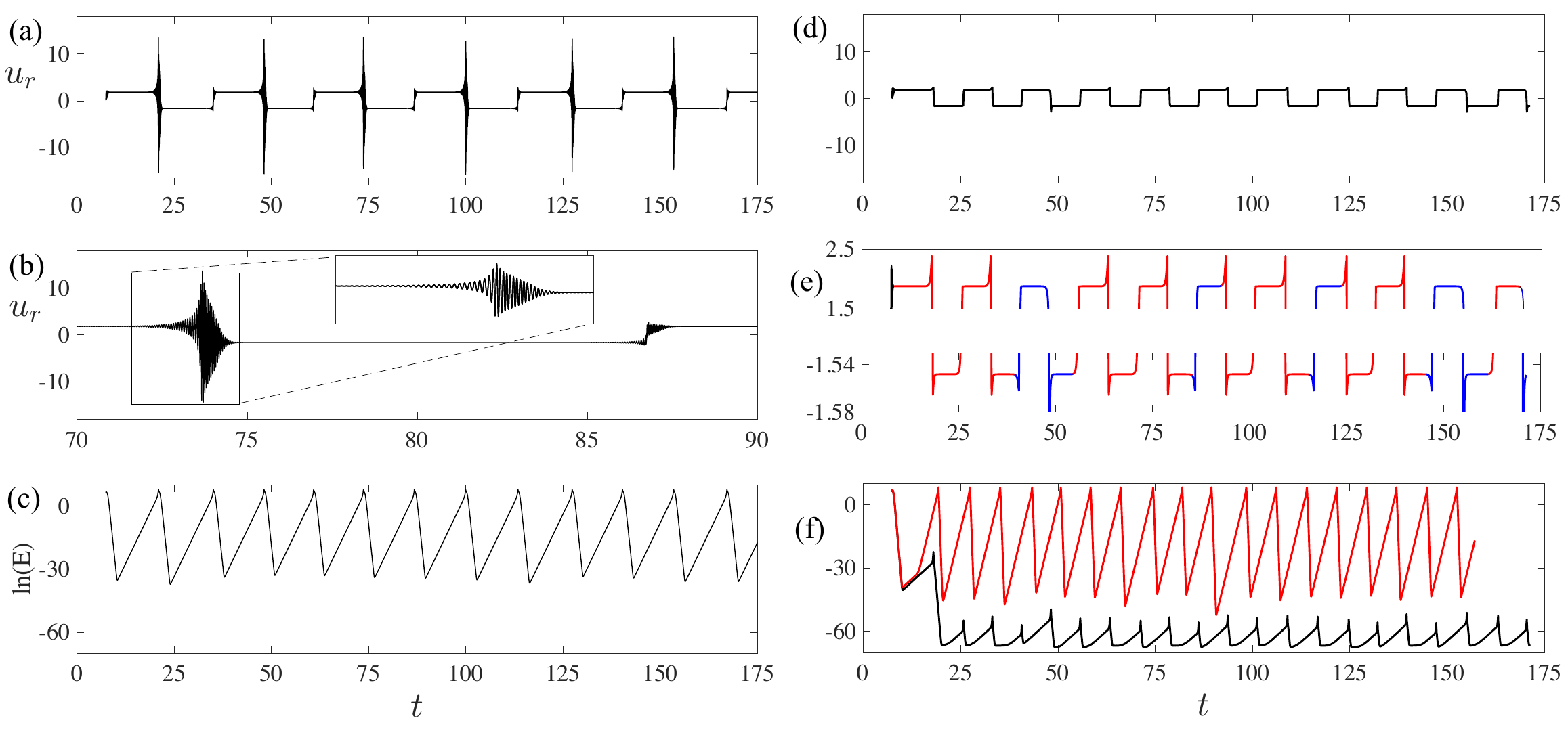}}
\caption{Timeseries for (a,b,c) oscillatory and non-axisymmetric heteroclinic cycle at $Re=-502$ and
for (d,e,f) non-oscillatory and axisymmetric heteroclinic cycle at $Re=-490$.  
(a,d) Complete timeseries alternating between two plateaus. (b,e) Enlargements of timeseries.
b) Enlargement of a single period from a non-axisymmetric cycle, with further enlargement around
the excursion. The differences between the growth/decay rates and frequencies
during the spiralling approach and departure are clear. e) Enlargements near the peaks
of the axisymmetric non-oscillatory excursions, highlighting their variation.
(c,f) Logarithmic plot of energy, which grows and decays exponentially.
The non-axisymmetric energy is shown in black and, for the axisymmetric heteroclinic cycle in (f),
the energy corresponding to axisymmetric but axially antisymmetric modes is shown in red.
Both types of heteroclinic cycles have the same plateaus, up to numerical roundoff.}
\label{fig:het_timeseries}
\vspace*{1cm}
\centerline{\includegraphics[width=\columnwidth]{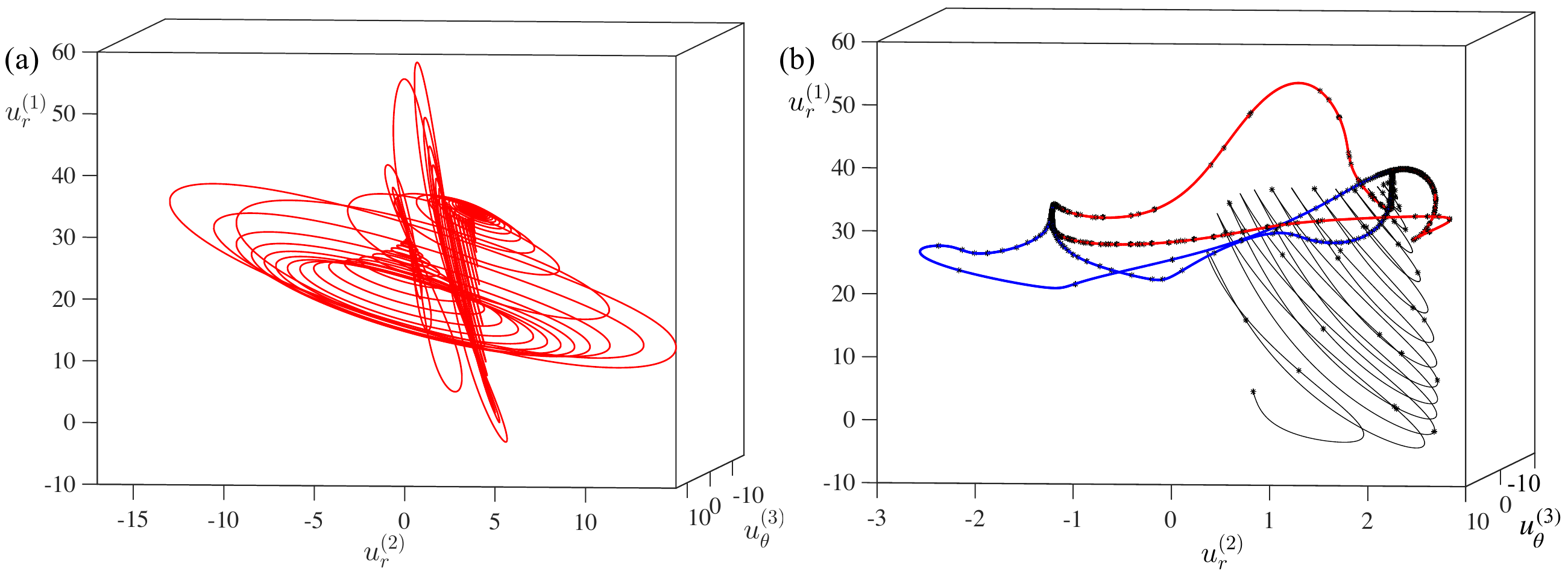}}
\caption{a) Phase portraits for a) oscillatory heteroclinic cycle at $\Rout=-502$
  and for b) non-oscillatory heteroclinic cycle at $\Rout=-490$ using
  $u_r(r=1.2,\theta=0,z=0)$, $u_r(r=1.2,\theta=\pi/3, z=\lambda_z/3)$, and
  $u_\theta(r=1.2,\theta=2\pi/3,z=2\lambda_z/3)$ corresponding to timeseries
  in figure \ref{fig:het_timeseries}. The trajectory in the oscillatory heteroclinic cycle
  spirals inwards along one curved surface and outwards along a perpendicular
  curved surface. Non-oscillatory phase portrait begins with a spiralling
  transient, then follows different paths into and out of the saddles. 
The color coding (blue and red) of these paths matches that in the timeseries,
where can be seen subtle differences between the excursions.
The saddles for the two cycles are the same; note the different scale 
for $u_r^{(2)}$ between (a) and (b).}
\label{fig:het_phase}
\end{figure}
\begin{figure}
\centerline{\includegraphics[width=0.9\columnwidth]{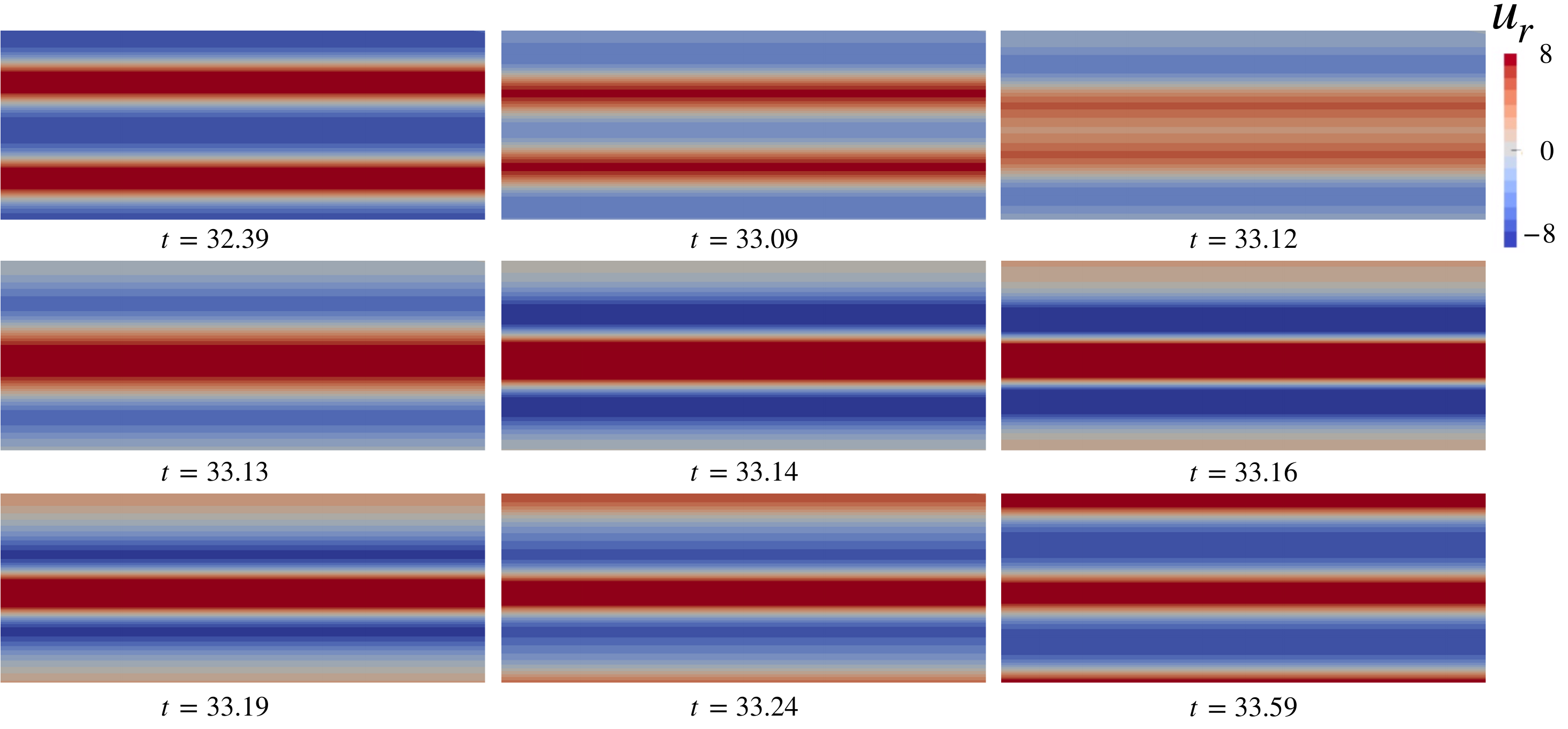}}
  \caption{Axisymmetric heteroclinic cycle at $\Rout=-490$.
Colors show the radial velocity in a $(\theta,z)$ slice at $r=1.26$.
The first (top left) and last (bottom right) visualizations show the 
saddles which anchor the heteroclinic cycles. 
These contain two pairs of axisymmetric vortices and differ by an 
axial phase shift of a single vortex.
For this heteroclinic cycle, the states remain axisymmetric 
throughout the rapid transition between the two saddles.}
\label{fig:Het_Snapshots_490}
\end{figure}
\begin{figure}
\centerline{\includegraphics[width=0.9\columnwidth]{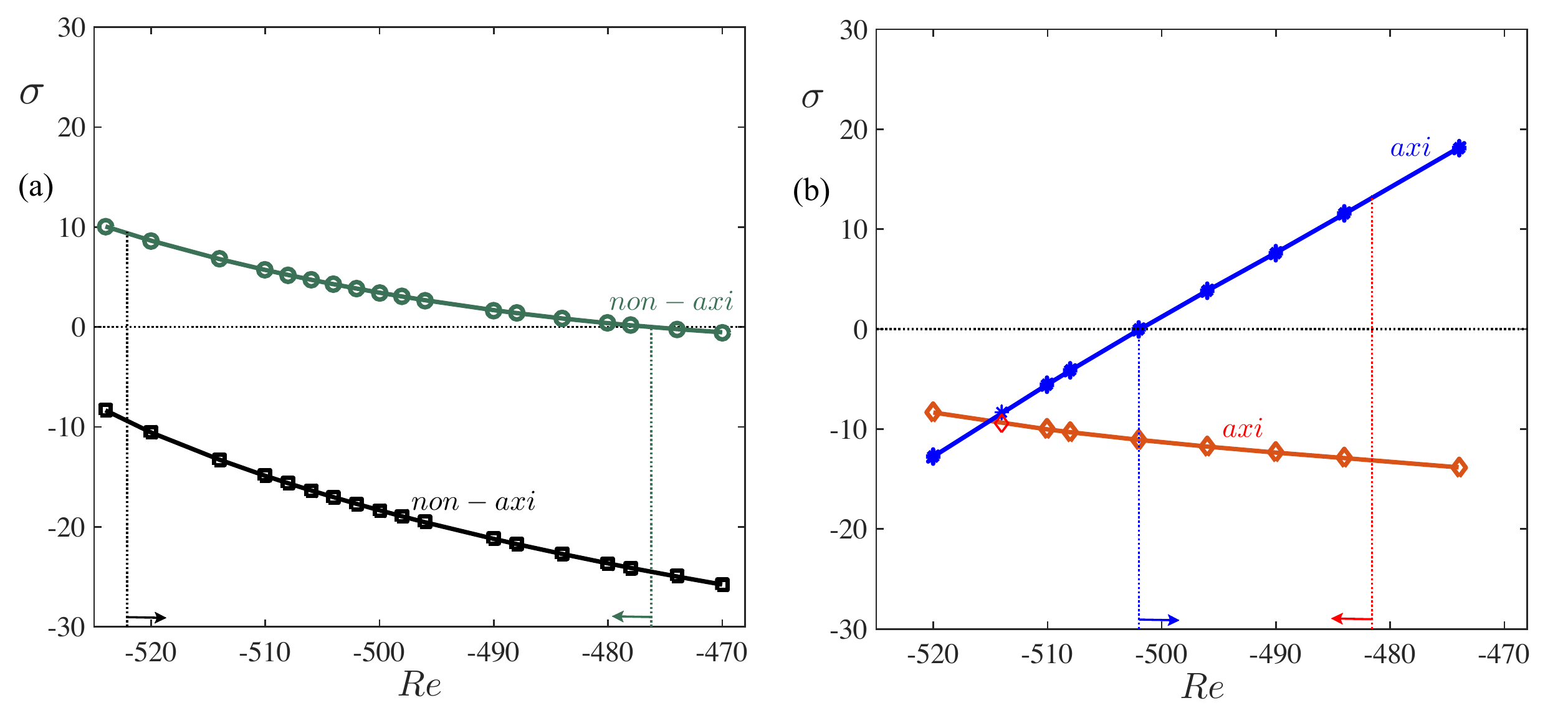}}
  \caption{Rate of approach to and departure from saddles 
    as a function of $\Rout$. These are the slopes of the
    logarithmic plots of figure \ref{fig:het_timeseries} and also 
the real parts of the leading eigenvalues.  a) 
    Non-axisymmetric oscillatory heteroclinic cycle.  The cycle 
exists when the eigenvalues are of opposite sign, here for $\Rout<-476$.
The cycle is stable within the space of these eigenmodes when the rate of approach exceeds the rate of departure, here for $\Rout>-522$.
b) Axisymmetric non-ocillatory heteroclinic cycle.  
The cycle exists when the eigenvalues are of opposite sign, here for $\Rout>-502$.
The cycle is stable within the space of these eigenmodes when the rate of approach exceeds the rate of departure, here for $\Rout<-482$.}
\label{fig:het_slopes}
\vspace*{1cm}
\centerline{\includegraphics[width=0.9\columnwidth]{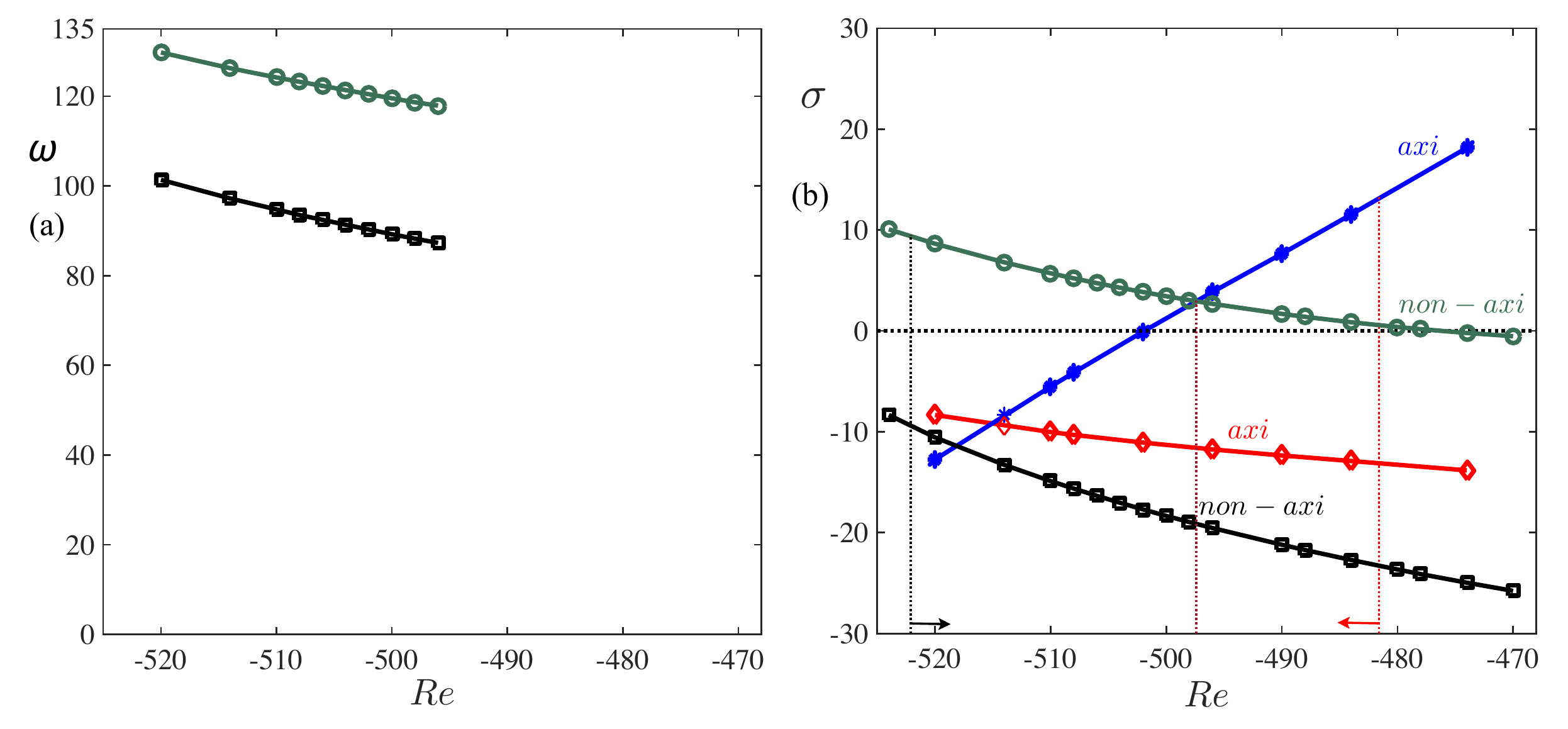}}
  \caption{a) Imaginary parts of eigenvalues involved in
    non-axisymmetric oscillatory heteroclinic cycle.  
    b) Real parts of eigenvalues involved in both heteroclinic cycles,
    axisymmetric and non-axisymmetric. Non-axisymmetric cycles are 
    observed for $\Rout < -496$, when the largest eigenvalue 
    (departure rate, dark green circles) corresponds to a non-axisymmetric eigenpair.
    Axisymmetric cycles are observed for $\Rout> -496$, when 
    the largest eigenvalue (blue stars) corresponds to an axisymmetric eigenvector.
    The other endpoints of the stability ranges, shown as the black rightwards
    and red leftwards arrows, involve the next-to-largest eigenvalues
    (rates of approach, red diamonds and black squares);
    see figure \ref{fig:het_slopes}
    and its caption. Summarizing, the stability range of the non-axisymmetric
    cycle is $\Rout\in[-522,-496]$, while that of the axisymmetric cycle is 
    $\Rout\in[-496,-482]$, which matches our observations.}
\label{fig:het_all}
\end{figure}

Pinter et al.~\cite{pinter2006competition,pinter2008bifurcation,pinter2008wave}
observed that the ribbon branch became unstable and 
was succeeded by oscillating cross-spirals as $\Rout$ was increased.
We observe this as well, near $\Rout\approx -525$.
As $\Rout$ is further increased, however, the oscillating cross-spirals are
themselves succeeded by a near-heteroclinic orbit, in which the system spends
the overwhelming majority of its time in one of two saddles containing
two pairs of axisymmetric vortices.
The two saddles differ by a quarter of an axial period (one vortex). The system passes
from one to the other via rapid non-axisymmetric excursions resembling the ribbon states.
Snapshots of the phases of an excursion taken from a heteroclinic cycle at $\Rout=-502$
are shown in figure \ref{fig:Het_Snapshots_502}.

Figure \ref{fig:het_timeseries}(a,b,c) shows timeseries from this heteroclinic orbit. Time
spent at the two saddles are seen as two plateaus. The approaches to
and departures from the saddles are oscillatory; this appears clearly
in the enlarged timeseries of figure \ref{fig:het_timeseries}b, as well as in the spirals of the phase
portrait of figure \ref{fig:het_phase}a. In the phase portrait, the two saddles are
distinguished, and the spiraling approaches to them are approximately
perpendicular to one another. In the timeseries of figure \ref{fig:het_timeseries}c, the
energy in the non-axisymmetric modes is seen to increase and to
decrease logarithmically.

As Re is further increased, this regime is succeeded by another type
of heteroclinic orbit, visualized in figure 11. Timeseries for this
cycle are shown in figure \ref{fig:het_timeseries}(d,e,f) for $Re = −490$. The two plateaus are
the same as for the previous heteroclinic cycle, but the approaches to
and departures from the saddles are not oscillatory. These approaches
and departures differ from excursion to excursion, as highlighted in
the enlargements in figure \ref{fig:het_timeseries}e. The difference between the excursions
is also seen in the phase portraits in figure \ref{fig:het_phase}b,
which begins with a transient oscillatory phase. The black dots in the
phase portrait are equally spaced in time and so accumulate on the
saddles. Trajectories are seen approaching and leaving the saddles along 
different paths at different angles.  Turning to the logarithmic
energy timeseries of figure \ref{fig:het_timeseries}(f), after an initial transient (corresponding to the
oscillatory transient in the phase portrait), the energy in the
non-axisymmetric modes becomes and remains very small. Instead, it is
the energy in the odd axial wavenumber modes which tracks the
heteroclinic cycle.  Since the saddles contain two stacked pairs of
vortices, their axial Fourier spectrum contain only even multiples of
the fundamental wavenumber $2\pi/\lambda_z$.  Excursions are
manifested by the logarithmic increase and decrease of the energy
contained in odd multiples of $2\pi/\lambda_z$ in the axial spectrum.
The variation from cycle to cycle is also seen in the logarithmic
timeseries.

The axisymmetric heteroclinic cycle is very similar to that observed
in the 1:2 mode-interaction
\cite{armbruster1988heteroclinic,nore2003}.  An erratic alternation
between a few different types of excursions is also seen in the cycles
of \cite{nore2003}.  As explained in these references, this cycle is a
consequence of the interaction of axial modes with wavenumbers
$2\pi/\lambda_z$ and $4\pi/\lambda_z$.  Furthermore, the heteroclinic
orbits can be characterized quantitatively by calculating the rates of
approach and departure from the saddles which anchor the heteroclinic
orbits.  These are the two leading eigenvalues (in
the axisymmetric subspace) of the saddles and are seen as the slopes
of the black curve in the logarithmic timeseries in figure
\ref{fig:het_timeseries}f.  We plot these as a function of $\Rout$ in
figure \ref{fig:het_slopes}b.  The existence of the heteroclinic cycle
requires eigenvalues of opposite signs; thus, the axisymmetric cycle
exists for $\Rout>-502$, where one of the leading eigenvalues becomes
positive while the other remains negative.  Stability of the cycle
requires that the rate of departure be no greater than the rate of
approach, i.e. the absolute value of the negative eigenvalue must
exceed the positive eigenvalue. This occurs for $\Rout<-482$.  Hence
the axisymmetric cycle exists and is stable within this axisymmetric
subspace in the range $\Rout\in[-502,-482]$.

The theory in \cite{armbruster1988heteroclinic,nore2003}
does not concern oscillatory heteroclinic cycles or complex eigenmodes.
However, assuming that the analysis of the oscillatory cycles is 
similar to that of the non-oscillatory cycles, we plot the 
slopes (or real parts of the eigenvalues) seen in the logarithmic
timeseries of figure \ref{fig:het_timeseries}c as a function of
$\Rout$.  We also plot the frequencies associated with
these eigenvalues in figure \ref{fig:het_all}a, 
associated with the oscillations in the 
timeseries in figure \ref{fig:het_timeseries}a,b and the spiraling in and out 
in the phase portraits of figure \ref{fig:het_phase}a.
Using the same criteria as for the non-oscillatory cycles, 
figure \ref{fig:het_slopes}a shows that the oscillatory heteroclinic cycle exists
and is stable within the subspace of these eigendirections for $\Rout\in[-522,-476]$.

The theory in \cite{nore2003,armbruster1988heteroclinic} also does not
address the relative stability of two kinds of heteroclinic cycles.
The transition that we observe when $\Rout$ is reduced from $-502$ to
$-490$ implies that the non-axisymmetric heteroclinic cycles is unstable
to the axisymmetric one at this value of $\Rout$.  In fact, we observe
non-axisymmetric heteroclinic cycles for $\Rout < -496$ and
axisymmetric cycles for $\Rout>-496$, separated by the point at which
axisymmetric and non-axisymmetric departure rates cross, as seen in
figure \ref{fig:het_all}b. This supports the idea that the cycle whose
departure rate from the saddles is smaller is unstable to the cycle
whose departure rate is greater \citep{ashwin1998attractors}.
Combining all of the criteria, the stability range of the non-axisymmetric
cycle is $\Rout\in[-522,-496]$, while that of the axisymmetric cycle is 
    $\Rout\in[-496,-482]$.

\section{Discussion}

The problem of characterizing the nonlinear frequency of a periodic flow
away from the threshold has led to one proposed solution: when a
linear stability analysis is carried out about the temporal mean, an
eigenvalue is obtained whose {\bf Real} part is {\bf Zero} and whose
{\bf Imaginary} part is the nonlinear {\bf Frequency}.  This RZIF
property has heretofore been studied for the cylinder wake, an open
cavity, and thermosolutal convection. It is strongly satisfied for the
cylinder wake and the traveling waves of thermosolutal convection with
oppositely directed density gradients, weakly satisfied for the open
cavity, and not at all satisfied for the standing waves of
thermosolutal convection.  Although the RZIF property is a natural
outcome of a near-monochromatic temporal spectrum, it is not entirely
clear when this occurs.

We have investigated RZIF in another configuration, counter-rotating
Taylor-Couette flow, which, like the thermosolutal convective case,
has a Hopf bifurcation leading to branches of traveling and standing
waves, here manifested as spirals and ribbons.  In the thermosolutal
case, the traveling waves display the RZIF property and have a
temporal Fourier spectrum which is highly peaked, while the standing
waves do not.  Here, both the spirals and the ribbons display the RZIF
property and both have temporal spectra which are equally
peaked. However, the ribbons are standing waves only in the axial
direction and are traveling waves in the azimuthal direction.  This
may be the cause of the difference between the Taylor-Couette and
thermosolutal cases.  The search for another counter-example to RZIF,
especially in a purely hydrodynamic flow without additional fields,
remains open.

During the course of our investigation, we discovered two heteroclinic
cycles, one non-axisymmetric and the other axisymmetric.  Both cycles
are anchored by the same saddles: axisymmetric states containing two
axially stacked pairs of vortices.  The axisymmetric heteroclinic
cycle is a manifestation of the classic 1:2 mode interaction, here the
interaction between axisymmetric states with one and two pairs of
vortices.  The non-axisymmetric heteroclinic cycle is not of this
type, since the approach to and from the saddles is oscillatory
via transients resembling ribbons.
However, it is anchored by the same axisymmetric saddles.  The
existence of two qualitatively different heteroclinic cycles
connecting the same saddles is an unusual and intriguing feature.
These cycles will be studied in more detail in a later investigation.

\begin{acknowledgments}
This work was performed in part using high performance computing resources
provided by the Institut du Developpement et des Ressources en
Informatique Scientifique (IDRIS) of the Centre National de la
Recherche Scientifique (CNRS), coordinated by GENCI (Grand Equipement
National de Calcul Intensif) through grant A0042A01119.
\end{acknowledgments}
%
\end{document}